\documentclass[12pt]{article}
\usepackage{pic02}
\usepackage{hyperref}
\usepackage{url}
\usepackage{graphicx}

\def\ifm#1{\relax\ifmmode#1\else$#1$\fi}
\def\DAF{DA\char8NE}  
\def\ffi{\ifm{\phi}}    
  \def\x{\ifm{\times}}
\def\gam{\ifm{\gamma}}  \def\pic{\ifm{\pi^+\pi^-}}
\def\pt#1,#2,{\ifm{#1\x10^{#2}}}   \def\epm{\ifm{e^+e^-}}
\renewcommand{\to}{\ensuremath{\rightarrow}}
\def\ao{\ifm{a_0}}  \def\po{\ifm{\pi^0}}  \def\et{\ifm{\eta}}

\def \a0{$a_0$}

\newcommand{\fog}{\ensuremath{f_0\gamma}}
\newcommand{\aog}{\ensuremath{a_0\gamma}}
\newcommand{\pipi}{\ensuremath{\pi\pi}}
\def \popo{$\pi^0\pi^0$}
\def \popog{$\pi^0\pi^0\gamma$}
\newcommand{\kakabar}{\ensuremath{K\bar{K}}}
\newcommand{\Jpsi}{\ensuremath{J/\psi}}
\renewcommand{\gg}{\ensuremath{\gamma\gamma}}
\def \qqbar{$q\overline{q}$}
\def \4q{$q\overline{q}q\overline{q}$}

\newcommand{\ks}{\mbox{$K_S$}}
\newcommand{\kl}{\mbox{$K_L$}}
\newcommand{\pip}{\mbox{$\pi^+$}}
\newcommand{\pim}{\mbox{$\pi^-$}}
\newcommand{\pio}{\mbox{$\pi^{0}$}}
\newcommand{\fo}{\ensuremath{f_0}}

\newcommand{\mpp}{\mbox{$M_{\pi\pi}$}}
\newcommand{\subfo}{\mbox{$_{f\!_{0}}$}}
\newcommand{\subao}{\mbox{$_{a\!_{0}}$}}
\newcommand{\phifog}{\mbox{$\phi\to f\!_{0}\gamma$}}
\newcommand{\phippg}{\mbox{$\phi\to\pio\pio\gamma$}}
\newcommand{\phiepg}{\mbox{$\phi\to\eta\pio\gamma$}}

\newcommand{\qqqq}{\mbox{$q\overline{q}q\overline{q}$}}
\newcommand{\lint}{\mbox{$L_{\rm int}$}}
\newcommand{\fietag}{\ensuremath{\phi\rightarrow\eta\gamma\;}}
\newcommand{\fietapg}{\ensuremath{\phi\rightarrow\eta'\gamma\;}}

\newcommand{\etap}{\ensuremath{\eta'\,}}
\newcommand{\Eta}{\ensuremath{\eta\,}}
\newcommand{\fip}{\ensuremath{\varphi_P\,}}
\newcommand{\tp}{\ensuremath{\vartheta_P \,}}
\newcommand{\ket}[1]{\ensuremath{\left|{#1}\right>}}
\begin{document}

\title{\bf Radiative $\phi$ decays.
\textit{}}
\author{
Antonella Antonelli        \\
{\em INFN, Laboratori Nazionali di Frascati}}
\maketitle

%
% photograph of author
%  This is where we will insert a photograph. To see what it would look like,
%  uncomment the following lines.
%
%\begin{figure}[htbp]
%\begin{center}
%
% include photograph for proceeding version
%
%\includegraphics[height=4.5cm]{Antonella_300.eps}
%
% insert a fixed vertical spacing instead for the ArXiv preprint
%
%\vspace{4.5cm}
%
%\end{center}
%\end{figure}

\baselineskip=14.5pt
\begin{abstract}
Radiative $\phi$ decays give us an excellent opportunity to study scalar and
pseudoscalar mesons below 1 GeV. In this paper, results from
different experiments are reviewed and compared.
\end{abstract}
\newpage

\baselineskip=17pt

\section{The puzzle regarding the scalar mesons \fo(980) and \ao(980)}
Although the existence and properties of the scalar mesons \fo(980) and 
\ao(980) are well established and have been known for about thirty years, 
the physical nature of these mesons is still unclear.
The first evidence for the \fo\ was in the reaction $p\pi \to n\pipi$.
The $I=J=0$ elastic \pipi\ cross section shows a dip close
to the \kakabar\ threshold. Similarly, in $p\pi \to n\kakabar$, a
sharp onset of inelasticity indicates
the presence of a dynamical structure strongly coupled to \kakabar. 
Evidence for strange quark content in the \fo\ is also confirmed by \Jpsi\ 
decays. In fact, the Mark-III, DM2, and BES collaborations measure 
BR(\Jpsi\to\fo$\phi$)$>$ BR(\Jpsi\to\fo$\omega$). However, the result 
obtained by the E791 collaboration \cite{E791_fo} shows that 56\% of the 
$D_s$ decay into 3$\pi$ proceeds via \fo, which suggests that \fo\ 
%is coupled to both $s\overline{s}$ and $n\overline{n}$.
has both $s\bar{s}$ and $n\bar{n}$ components.
The total width of the \fo\ meson has been measured by several experiments and
ranges between 40--100 MeV; moreover, the \fo\ is weakly coupled to 
photons [$\Gamma(\fo\to\gg)\approx 0.3$ KeV].
The first evidence for the \ao\ was found in the $\eta\pi$ system in
$Kp \to \Sigma\eta\pi$. Today, there are very accurate data from the 
Crystal Barrel, GAMS, Obelix and E852 collaborations \cite{BNL852,crystal}. 
As in the case of the \fo,
the \ao\ also has a small width compared to ordinary mesons with the same 
mass, and a small \gg\ coupling.
In short, these two scalar mesons are very close in mass, and 
have small widths and small \gg\ couplings. The standard interpretations of 
these states as $q\bar{q}$ mesons is not favoured. The total widths 
are much smaller than the $\sim$500 MeV expected from the $q\bar{q}$ 
prediction, the \gg\ partial widths measured are a factor 10--20 smaller than
the $q\bar{q}$ prediction, and the \kakabar\ coupling is too large
for an OZI-forbidden decay. The theoretical situation is further complicated 
because there are many scalars below 1.5 GeV. 
Many different interpretations of these states have been proposed: 
\kakabar\ molecules \cite{kk}, four-quark states \cite{4quarks}, and, in the case of the \fo, 
glueballs. Interest in studying light scalar mesons
also comes from an old suggestion by Gribov that foresees the existence of
a peculiar state with the vacuum quantum numbers and a mass close to the
proton mass to explain quark confinement. Most recently, Close and T\"{o}rnqvist
\cite{closetorquist} have interpreted these mesons as the Higgs 
nonet of a hidden U(3) symmetry.
From all of the above considerations, it is obviously urgent to clarify the
situation. The radiative $\phi$ decays are very useful for addressing this
puzzle. In fact, the absolute rates for $\phi$\to\fog\ and \aog\ and the 
mass spectra of the \fo\ and \ao\ are very senvitive to the nature of 
these scalar particles, as shown in Tab.~\ref{Tab:comp}.

\begin{table}[htbp]
\centering
\caption{\it Branching ratio predictions for different models.}
\label{Tab:comp}
\vskip 0.1 in
\newcommand{\m}{\hphantom{$0$}}
\renewcommand{\tabcolsep}{0.6pc} % enlarge column spacing
\renewcommand{\arraystretch}{1.2} % enlarge line spacing
\begin{tabular}{|l|c|c|c|c|} \hline
Channel                  &  \qqbar      &  \4q       & \kakabar  \\ \hline
\hline
BR($\phi$\to\fog)      &  $5\times10^{-5}$ &  $10^{-4}$ &  $10^{-5}$   \\
BR($\phi$\to\aog)      &  $10^{-5}$   &  $10^{-4}$ &  $10^{-5}$    \\ \hline
\end{tabular}
\end{table}

\section{Radiative $\phi$ decays to \fo, \ao}
Most of the data on the $\phi$\to\fog\ and \ffi\to\aog\ decays come from 
three
experiments: KLOE \cite{kloefo,kloeao} at the Frascati $\phi$-factory \DAF \cite{dafne}, 
and SND \cite{sndfo,sndao} and CMD-2 \cite{cmd2fo,cmd2ao} at VEPP-2M in
Novosibirsk. The published analyses are based on $5.3\times10^7$ $\phi$ decays
for the KLOE experiment, with large acceptances, and $2\times10^7$ $\phi$ 
decays for the VEPP-2M experiments, with smaller acceptances.
The \fo\ has been studied via its decay to \popo.
The \ao\ has been studied via its decay to $\eta\pio$, 
with $\eta$\to$\gamma\gamma$ and $\eta$\to\pip\pim\pio. 
For this last channel, only data from the KLOE experiment are 
available.

\subsection{{$\phi$\to\popog}}

Two amplitudes contribute to the \phippg\ final state: 
$\ffi \to S \gam; S \to \po\po$ and $\ffi \to \rho^0 \po; \rho^0 
\to \po \gam$,
where $S$ is a scalar meson. The main backgrounds come from 
$\epm\to\omega\po\to\po\po\gam$, $\ffi\to\et\po\gam\to5\gam$,
$\ffi\to\et\gam\to3\po\gam$ with two undetected photons, and
$\ffi\to\et\gam\to3\gam$ with two additional photons from accidentals. 

The natural cross sections for the signal and background channels are listed
in Tab.~\ref{Tab:Bckg}. The $\omega\pi$ background is the most important,
since the background from $\eta\gamma$ into 3- and 7-photon final states is
immediately reduced by exploiting the detector hermiticity and the photon 
timing.

\begin{table}[htbp]
\centering
\caption{\it Cross sections for signal and background channels for \phippg.}
\label{Tab:Bckg}
\vskip 0.1 in
\newcommand{\m}{\hphantom{$0$}}
\renewcommand{\tabcolsep}{0.6pc} % enlarge column spacing
\renewcommand{\arraystretch}{1.2} % enlarge line spacing
\begin{tabular}{|l|c|c|} \hline
Channel          &  cross section (nb)    \\ \hline
\hline
\fog             &  $\approx$ 0.4      \\
$\omega\pi$      &  $\approx$ 0.5       \\
$\eta\pi\gamma$  &  $\approx$ 0.1       \\
$\eta\gamma$     &  $\approx$ 17.0   \\ \hline
\end{tabular}
\end{table}
\par

\begin{figure}[htbp]
\includegraphics[width=13cm]{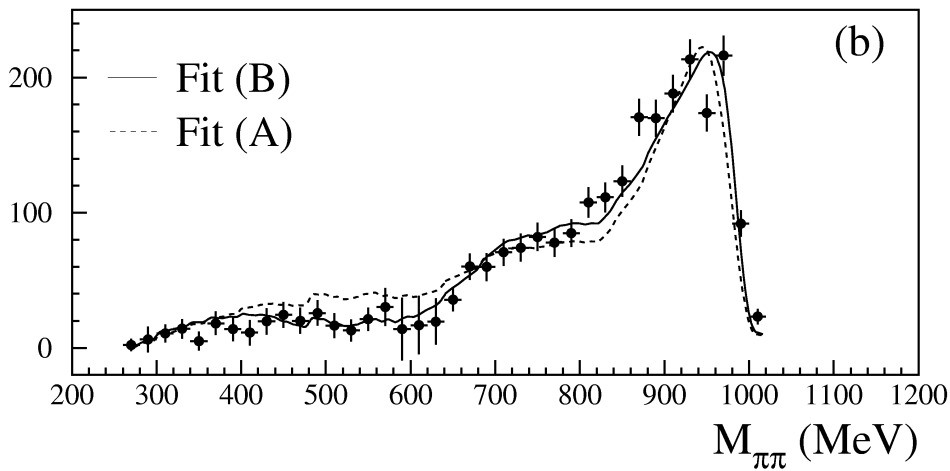}
 \caption{\it
     \pipi\ invariant mass spectrum, KLOE experiment.
    \label{kloe} }
\end{figure}

\begin{figure}[htbp]
  \centerline{\hbox{ \hspace{0.2cm}
    \includegraphics[width=6.0cm]{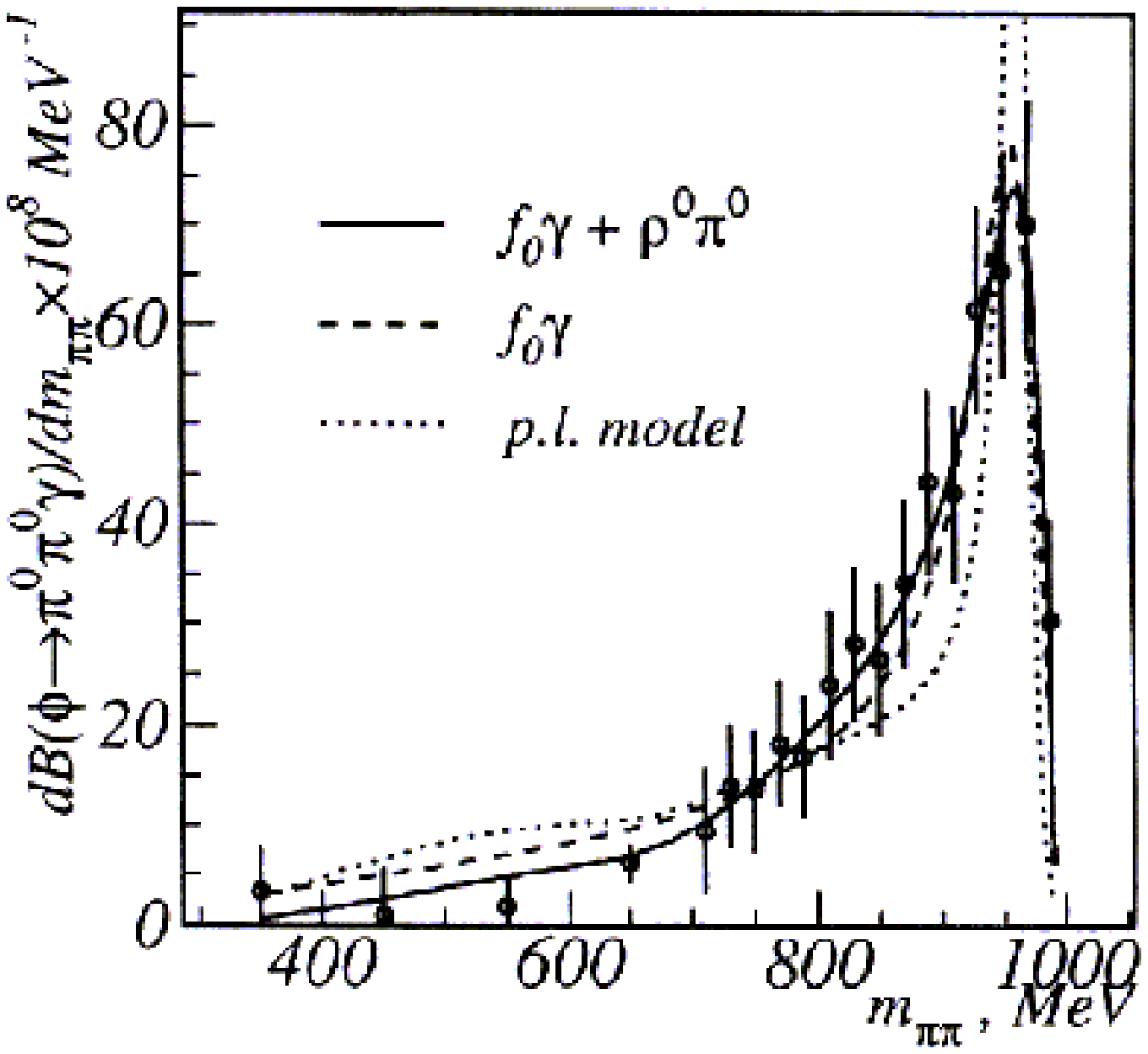}
    \hspace{0.6cm}
    \includegraphics[width=6.0cm]{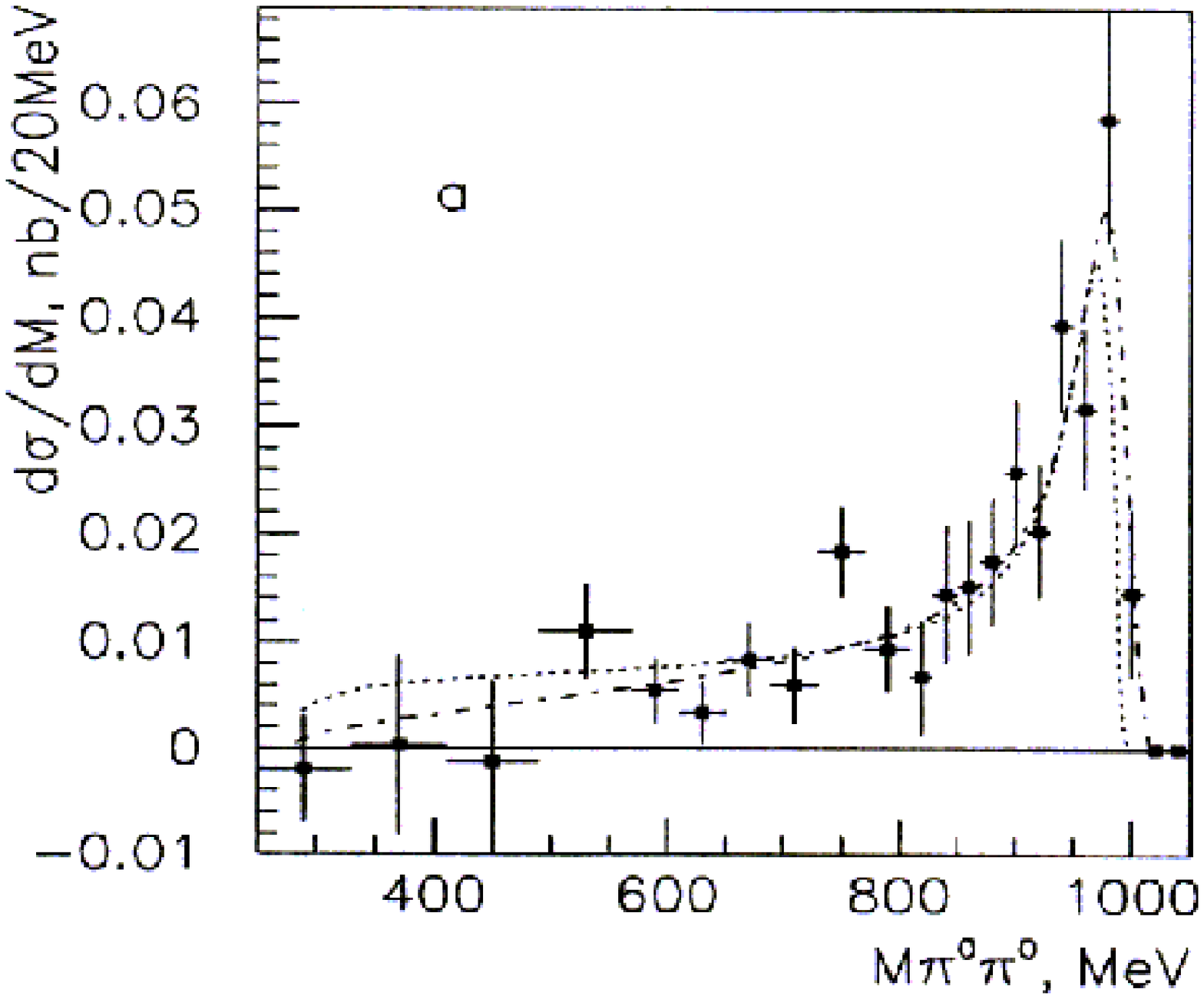}
    }
  }
 \caption{\it
      \pipi\ invariant mass spectra, CMD-2 (left) and SND (right) experiments.
    \label{sndcmd2} }
\end{figure}

All three collaborations have performed similar analyses. The selection of 
\phippg\ events starts from the five-photon sample. A cut on the total energy
in the calorimeter rejects the \ffi\to\ks\kl\ background. The photons are
paired to search for a \pio\ and a cut on $|M_{\pi\gamma}-M_{\omega}|$ is
performed to veto the $\omega\pi$ background. A kinematic fit  
requiring the \pio\ masses further reduces the background. The numbers of 
signal events and the mean efficiencies 
are shown in Tab.~\ref{Tab:statf0} for each of the three collaborations.
\begin{table}[htbp]
\centering
\caption{\it Signal statistics and efficiencies for \phippg.}
\vskip 0.1 in
\label{Tab:statf0}
\newcommand{\m}{\hphantom{$0$}}
\renewcommand{\tabcolsep}{0.6pc} % enlarge column spacing
\renewcommand{\arraystretch}{1.2} % enlarge line spacing
\begin{tabular}{|l|c|c|} \hline
EXPERIMENT       & Signal events & Mean efficiency ($\%$) \\ \hline\hline
KLOE             & $2438 \pm 61$ & $\sim40$     \\
SND              & $419 \pm 31$  & $\sim20$     \\
CMD-2            & $268 \pm 27$  & $\sim12$     \\  \hline
\end{tabular}
\end{table}
The \pipi\ invariant mass spectra are shown in Figs.\ref{kloe} and \ref{sndcmd2}. 
A clear peak in the \fo\ region is seen.

\subsection{Model for fits to \mpp\ spectrum}
All the three experiments have fit the mass spectrum using the same model.
This spectrum is taken as the sum of $S\gamma$, $\rho\pi$, and 
interference terms: $f(m)=f_{S\gamma}(m) + f_{\rho\pi}(m) + f_{\rm int}(m)$.
For the scalar term, the kaon loop model is used \cite{AchIvanch}. In
this model, the radiative \ffi\ decays to a scalar proceed through a charged
kaon loop, and the scalar term can be written as: 
\begin{equation}
 f_{S\gamma}(m) = \frac{2\,m^2}{\pi}\:
 \frac{\Gamma_{\phi S\gamma}\Gamma_{S\pi^{0}\pi^{0}} }{|D_S|^2}\:
 \frac{1}{\Gamma_\phi}.
\end{equation}
\begin{equation}
\Gamma_{\phi f\!_{0}\gamma} (m) =
\frac{g^2_{f\!_0K^+K^-}g^2_{\phi K^+K^-}}{12\pi}\:
\frac{|g(m)|^2}{M_{\phi}^2}\:
\left( \frac{M^2_{\phi}-m^2}{2M_{\phi}} \right),
\end{equation}
where $g_{\phi K^+K^-}$ and $g_{f\!_{0} K^+K^-}$ are the couplings
and $g(m)$ is the loop integral function.
For the inverse propagator, $D_S$, the finite width 
corrections \cite{AchIvanch} are taken into account. 
The parametrization of Ref.~\cite{AchInt} has been used for the 
$\rho\pi$ and interference terms.
A recent measurement \cite{E791_sigma,E791_sigma1}
reports the existence of a 
scalar $\sigma$ with $M_\sigma = (478^{+24}_{-23}\pm 17)$ MeV and 
$\Gamma_\sigma = (324^{+42}_{-40}\pm 21)$ MeV.
The contribution of this meson \cite{Ankara} has also been included in 
fits to the mass spectrum, giving different results as discussed below. 
\par
The observed mass spectrum, $S_{\rm obs}(\mpp)$, is fit by folding 
the experimental efficiency and resolution into the theoretical shape 
after properly normalizing for the cross section for $\phi$ production, 
$\sigma(\phi)$, and the integrated luminosity, \lint. 
The KLOE collaboration has performed a fit including $\sigma$ and $\rho\pi$ 
contributions. They find that the data prefer a negligible $\rho\pi$
contribution and a negative interference between the \fo\ and $\sigma$ 
amplitudes at $\mpp < 700$ MeV. The CMD-2 collaboration has done a similar 
fit and they find a negligible $\sigma$ contribution. 
SND has fit the data considering the \fo\ contribution only. It is worth 
noticing that, due to the lack of statistics, the VEPP-2M exeriments
do not have much sensitivity in the  $\sigma$ region.
The values of the coupling constant obtained by KLOE \cite{kloefo}, SND
\cite{sndfo}, and CMD-2 \cite{cmd2fo}
are listed in Tab.~\ref{Tab:FitRes}, and are compared with the results from 
the WA102 and E791 collaborations.
The coupling constants obtained from $\phi$ decays agree with each other, and 
differ from the WA102 result on \fo\ production in central $pp$ collisions 
\cite{WA102} and from results obtained when the \fo\ is produced in 
$D_s^+\to\pip\pim\pip$ decays \cite{E791_fo}, in which $g_K$ 
is consistent with zero. In the same table, measurements of BR(\phifog) 
for \mpp $>$ 700 MeV are also listed. 

\begin{table}[htbp]
\caption{\it Comparison of \fo\ parameters from different experiments.}
\vskip 0.1 in
\label{Tab:FitRes}
\renewcommand{\tabcolsep}{0.8pc} % enlarge column spacing
\renewcommand{\arraystretch}{1.2} % enlarge line spacing
\begin{scriptsize}
\begin{tabular}{@{}|l|c|c|c|c|c|} \hline
                            &  KLOE        &  SND          & CMD-2          & WA102       &  E791
 \\ \hline\hline 
$M\!\subfo$ (MeV)           & $973\pm 1$   &  $969 \pm 5$  &  $975  \pm 7$  & $987 \pm 8$ &  $977  \pm 4$   \\
$g^2_{f\!_0 K^+K^-}/(4\pi)$ (GeV$^2$)       &  $2.79 \pm 0.12$    &  $2.47 \pm 0.73$ &
  $1.48 \pm 0.32$  &  $0.40\pm 0.06$ &  $0.02 \pm 0.05$ \\
$g^2_{f\!_0 K^+K^-}/g^2_{f\!_0\pi^+\pi^-}$  & $4.00 \pm 0.14$    & $4.6 \pm 0.8$  
 & $3.61 \pm 0.62$  & $1.63 \pm 0.46$   & ---   \\
$g_{\phi\sigma\gamma}$      &  $0.060\pm 0.008$  & --- & --- & --- & --- \\ 
\parbox{1.5in}{BR(\phippg)$\times10^4$\\\mpp $>$ 700 MeV}  & $0.96 \pm 0.05$ & $1.03 \pm 0.09$ &
 $0.92 \pm 0.09$ &---&---\\
\hline
\end{tabular}
\end{scriptsize}
\end{table}

\subsection{{\phiepg}}
Production of the \ao~meson followed by $\ao\to\et\po$ 
dominates the final state \et\po\gam\ in \ffi-decays. A small 
contribution from $\ffi \to \rho^0\po$, $\rho \to \et\gam$ is present.
\par
The \phiepg\ decay is studied by KLOE \cite{kloeao}, CMD-2 \cite{cmd2ao} and
SND \cite{sndao} using the 
\et\to\gam\gam\ decay mode; the analysis starts from the same 
$5\gam$ sample as for the \fo\ selection. 
The main backgrounds come from the $\phi\to\pi^0\pi^0\gamma$ channel, 
which is dominated by $\ffi\to\fog$,
the non-resonant $e^+e^-\to\omega\pi^0$ interaction with
    $\omega\to\pi^0\gamma$, and $\ffi\to\et\gam$ with $\et\to\gam\gam$
and  $\eta\to\pi^0\pi^0\pi^0$.
The numbers of signal events and the mean efficiencies are summarized in Tab.\ref{bckg1}.
\begin{table}[htbp]
\centering
\caption{\it Signal statistics and efficiencies for \phiepg.}
\vskip 0.1 in
\label{bckg1}
\newcommand{\m}{\hphantom{$0$}}
\renewcommand{\tabcolsep}{0.6pc} % enlarge column spacing
\renewcommand{\arraystretch}{1.2} % enlarge line spacing
\begin{tabular}{|l|c|c|} \hline
EXPERIMENT       & Signal events & Mean efficiency ($\%$)  \\ \hline
\hline
KLOE             & $607\pm36$ & 33  \\
SND              & $35\pm6$   & 2.3 \\
CMD-2            & $80\pm22$  & 4   \\  \hline
\end{tabular}
\end{table}
The KLOE collaboration also makes use of the \phiepg, 
$\eta\to\pim\pip\pio$ channel.
In this case, the lower statistics are compensated for by the fact that there 
are no backgrounds with the same final state. The KLOE collaboration selects 
197 events with an estimated background of $4 \pm 4$ events and an efficiency
of $\approx 19\%$.
\par
The fit to the mass spectrum for the \ao\ is performed using the kaon loop
model as in the case of the \fo. The $\ffi\to\rho\po$, $\rho\to\et\gam$
contribution is also considered by the KLOE collaboration.
The SND collaboration fits the data assuming only the \ao\ contribution.
The KLOE collaboration performs a combined fit using the two $\eta$ decay
modes and setting $M_{\ao}$=984.8 MeV, from 
Ref.~\cite{PDG}. The free parameters of the fit are
the branching ratio
for the $\ffi\to\rho\po$ contribution and the two coupling constants.

\begin{figure}[htbp]
\begin{center}
\includegraphics[width=8cm]{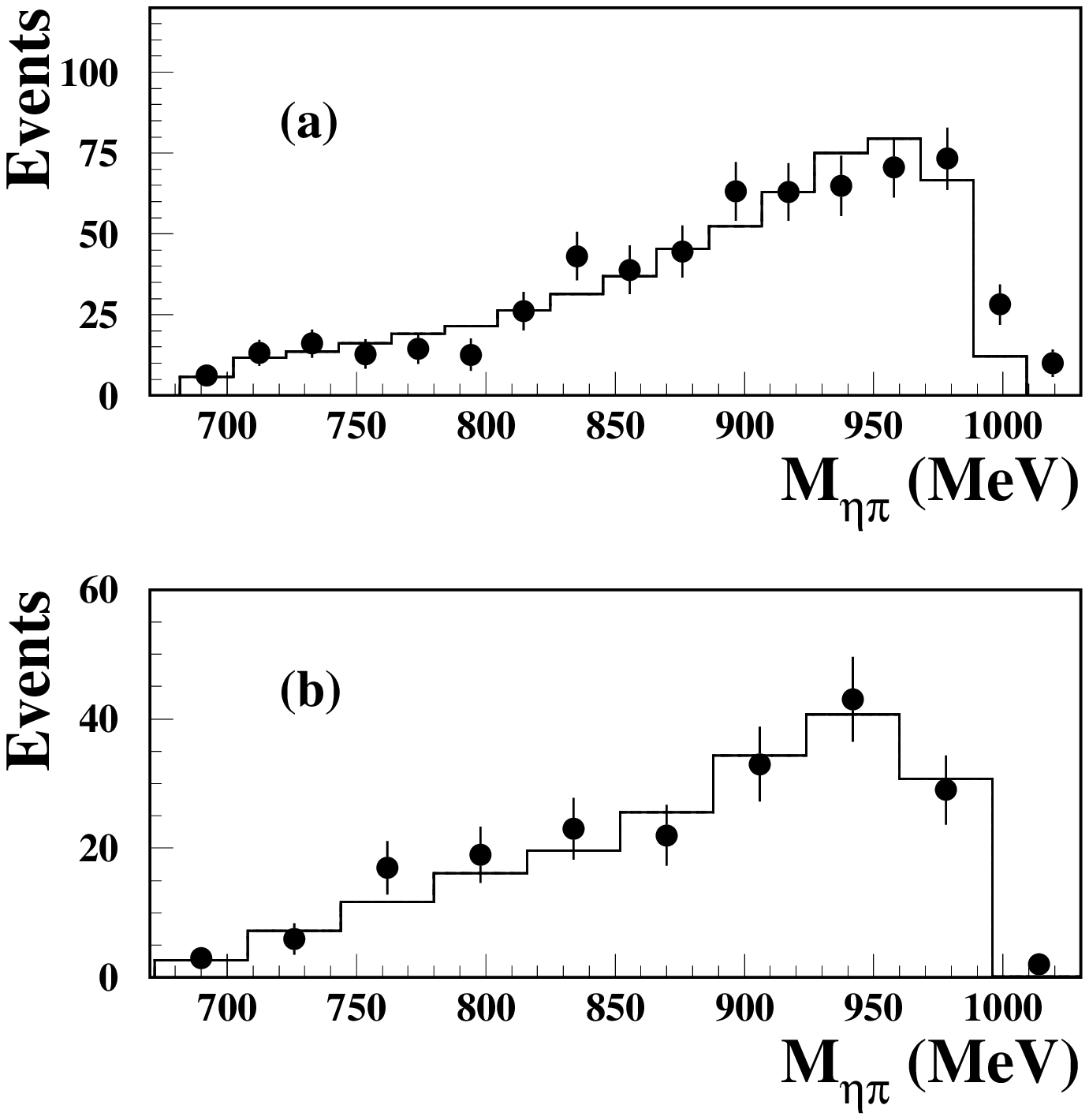}
\end{center}
 \caption{\it KLOE result for combined fit: comparison of data
(exp. points) vs. fit (histogram) for
  \ffi\to\et\po\gam\ with (a) \et\to\gam\gam\ and (b) \et\to\pic\po.}
    \label{kloefita0} 
\end{figure}

The CMD-2 collaboration does not attempt to fit the mass
spectrum due to lack of statistics; only BR(\phiepg) is quoted.
In Fig.\ref{kloefita0}, 
the mass spectrum from the KLOE experiment is shown for the two \et\ decay 
channels.
The values of the coupling constant obtained by KLOE, SND, and CMD-2
are listed in Tab.~\ref{Tab:FitResao} and are compared with the results from 
BNL-E852 \cite{BNL852} and with the various Crystal Barrel results 
\cite{crystal}

\begin{table}[htbp]
\caption{\it Comparison of \ao\ parameters from different experiments.}
\label{Tab:FitResao}
\vskip 0.1 in
\newcommand{\m}{\hphantom{$-$}}
\newcommand{\cc}[1]{\multicolumn{1}{c}{#1}}
\renewcommand{\tabcolsep}{0.8pc} % enlarge column spacing
\renewcommand{\arraystretch}{1.2} % enlarge line spacing
\begin{footnotesize}
\begin{tabular}{@{}|l|c|c|c|c|c|} \hline
                            &  KLOE        &  SND  & CMD-2 & E852&  \parbox[b]{0.75in}{\center Crystal\\Barrel} \\ \hline
\hline 
$M\!\subao$ (MeV)           & \parbox[b]{0.75in}{\center 984.8\\(fixed)} &  $995^{+52}_{-10}$  &  ---
 &$991  \pm 3$ &  $1000  \pm 2$   \\
$g^2_{a\!_0 K^+K^-}/(4\pi)$ (GeV$^2$)       &  $0.40 \pm 0.04$    & $1.4^{+9.4}_{-0.9}$    &
  &---  &  --- \\
$g_{a\!_0\eta\pi}/g_{f\!_0 K^+K^-}$  & $1.35 \pm 0.09$    & $0.75 \pm 0.52$  
 & ---  & $1.05 \pm 0.06$   &0.93 -- 1.07   \\
 BR(\phiepg)$\times10^5$   & $7.4 \pm 0.7$ & $8.8 \pm 1.7$ &
 $9.2 \pm 2.6$ &---&---\\
\hline
\end{tabular}
\end{footnotesize}
\end{table}

%\begin{figure}[htbp][htb]
%\includegraphics[width=6cm]{sndfita0.eps}
% \caption{\it SND fit result. Points are data; line shows fit
%   considering \ao\ contribution only.}
%    \label{sndfita0} 
%\end{figure}

\subsection{Comparison with theoretical expectations}
The measured coupling constants for the \ao~and \fo~can be compared 
with the predictions for the $q\bar{q}$ and $qq\bar{q}\bar{q}$ models. 
These two models make predictions for the ratio of the coupling constants, $R$,
based on SU(3) algebra assuming no OZI-violation. Qualitative predictions 
are available for the absolute values of the couplings. Tab.~\ref{conf_model} 
shows these predictions compared with KLOE results.
\begin{table}[htbp]
\caption{\it Comparison of KLOE results with predictions
  from $q\bar{q}$ and $qq\bar{q}\bar{q}$
 models. $q\bar{q}$ refers to either
 $\fo=(u\bar{u}+d\bar{d})/\sqrt{2}$ (b) or $\fo=s\bar{s}$ (a).
 Here, $R\subfo$ is $g^2_{f\!_0 K^+K^-}/g^2_{f\!_0\pi^+\pi^-}$ and 
 $R_{a_0}$ is $g^2_{a\!_0 K^+K^-}/g^2_{a\!_0\eta\pi}$.
 }
\vskip 0.1 in
  \begin{center}
\begin{tabular}{|c|c|c|c|c|} \hline
 Experiment & $g^2_{f\!_0K\bar{K}}/4\pi$ & $g^2_{a_0K\bar{K}}/4\pi$ & $R\subfo$ & $R_{a_0}$ \\
\hline
KLOE & $2.79\pm0.12$ & $0.40\pm0.04$ & $4.00\pm0.14$ & $0.55\pm0.07$ \\
$qq\bar{q}\bar{q}$ &''superallowed'' & ``superallowed'' & $4\div8$ &
1.2 \\
$q\bar{q}$ (a)  & ``OZI-allowed'' & ``OZI-forbidden'' & 4 & 0.43 \\
$q\bar{q}$ (b)  & ``OZI-forbidden'' & ``OZI-forbidden'' & 0.5 & 0.43 \\
\hline
\end{tabular}
  \end{center}
\label{conf_model}
\end{table}
The predictions have been evaluated using a value of $\theta_{P}=-13.7^\circ$ 
for the pseudoscalar mixing angle.
From this comparison, it is evident that:
\begin{enumerate}
\item The \fo\ parameters are not compatible with
    the predictions for the $q\bar{q}=u\bar{u}+d\bar{d}$ model.
    $R\subfo$ and $g_{f\!_0K\bar{K}}$ are both too large.
\item $R\subfo$ is compatible both with \qqqq\ model parameters 
    and with the expectation for $s\bar{s}$. 
\item The $a_0$ coupling, $g_{a_0K\bar{K}}$, is too small with respect 
    to predictions for the \qqqq\ model. 
\item $R_{a_0}$ seems to indicate better agreement with $q\bar{q}$ model.
\end{enumerate}   
Predictions for the values of the branching ratios have also been made in the
framework of the linear sigma model \cite{Escri,Bramon:2002iw} and in the unitarized
chiral model approach \cite{Marco}, and show rather good agreement.

\section{Pseudoscalar:~\fietag, \fietapg}
Radiative decays of light vector mesons to pseudoscalars have been used 
as a testing ground since the early days of the quark model \cite{BeMor65}.
The branching ratio of the decay \fietapg is particularly interesting 
since its value can probe the $\left|s\bar{s}\right>$ and gluonium content
of the \etap \cite{Close92}. In particular, the ratio $R$ of the branching 
ratios for \fietapg and \fietag can be related to the \Eta-\etap mixing 
parameters \cite{Ros83,BraEsSca97,BraEsSca99,BraEsSca01,Feld00} and 
determine the mixing angle in the flavor basis \fip, which has been
identified as the best suited parameter for a process-independent
description of the mixing. 
In fact, within the two-mixing-angles scenario, which has emerged from
an extended chiral perturbation theory framework \cite{KaisLeut98} as well 
as from phenomenological analyses \cite{EsFre99}, it has been demonstrated 
that the two mixing parameters in the flavor basis are equal, apart from 
terms which violate the OZI rule \cite{FeldKroll98,Defazio00}. It is 
therefore safe to use a single mixing angle in this basis. 
The measurements available to date on BR(\fietapg) come from the KLOE \cite{kloeetap}, CMD-2\cite{CMD200b},
and SND\cite{SND99} collaborations. The most precise measurements come from the
KLOE collaboration. 
\par
KLOE has analyzed a sample of $5.3\times10^7$ $\phi$
decays, looking for \fietapg via the $\etap\to\pip\pim\eta$ and $\fietag$,
$\eta\rightarrow\pip\pim\pio$ decay chains. For both decay chains, the
$\pipi3\gam$ final state is used; therefore, many common systematic effects
approximately cancel out in the ratio $R= BR(\fietapg)/BR(\fietag)$.
After the analysis, KLOE finds 128 \etap events with an efficiency of
$\approx 23 \%$ and a background level of 6$\%$.
The value $R = \left(4.7 \pm 0.5 \pm 0.3 \right)\cdot
10^{-3}$ has been obtained.

\begin{figure}[htbp]
  \centerline{\hbox{ \hspace{0.2cm}
      \includegraphics[width=6cm]{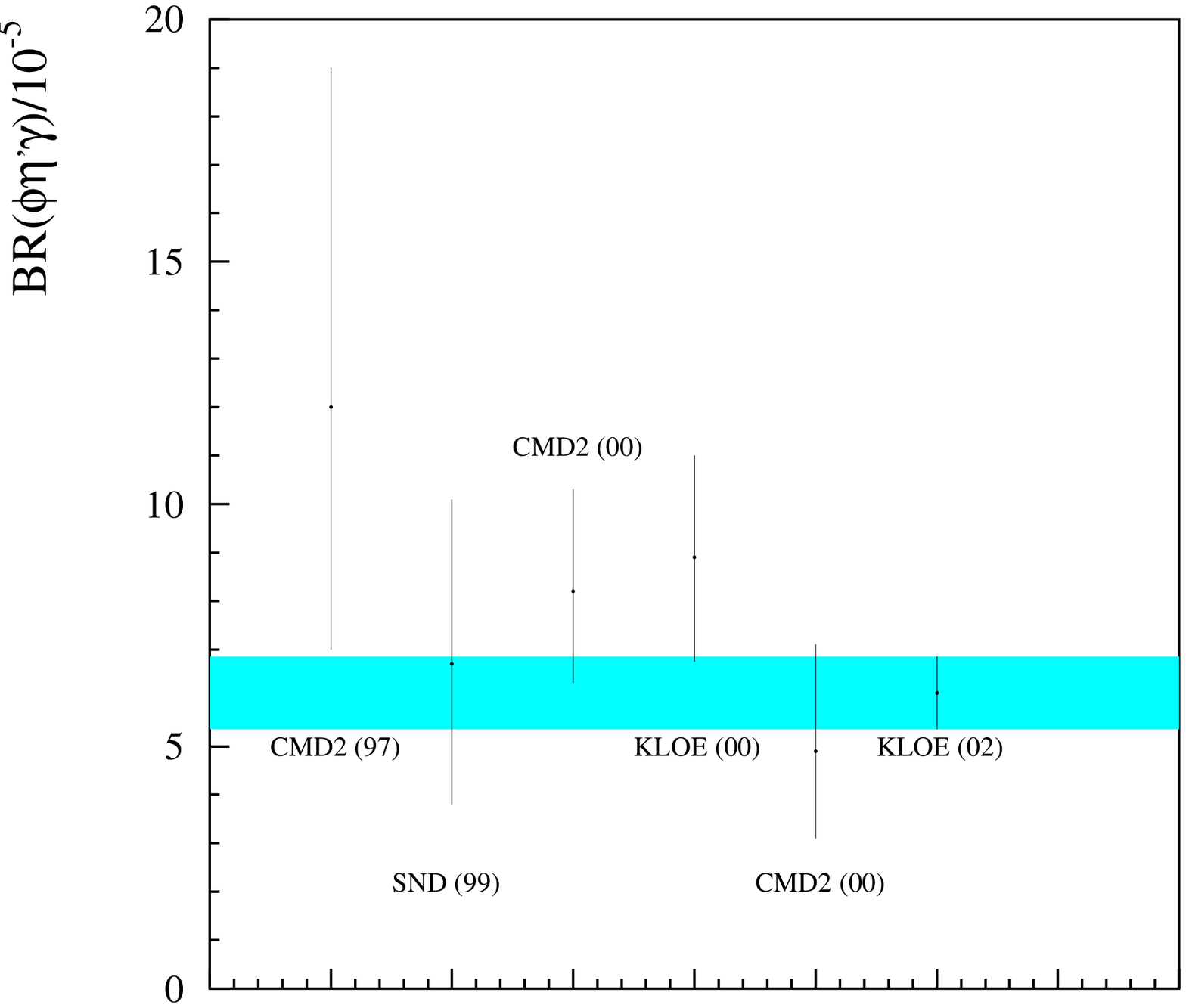}
      \hspace{0.6cm}
      \includegraphics[width=6cm]{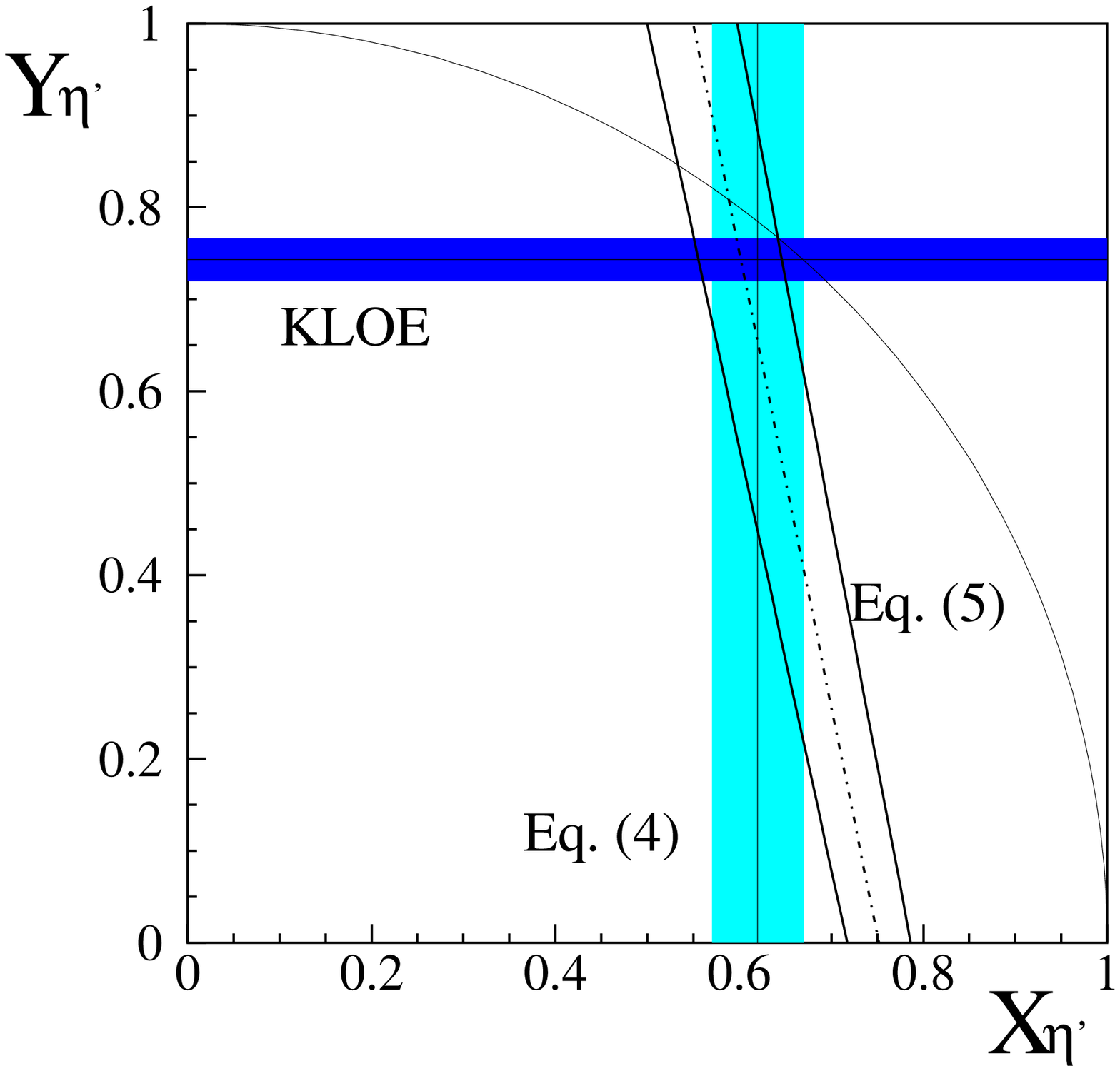}
      }
  }
\caption{\it Left: Determinations of the BR(\fietapg) in the literature:
CMD2\cite{CMD297,CMD200,CMD200b}; SND\cite{SND99}; 
KLOE\cite{kloeetap}. Right: Bounds on $X_{\etap}$ and $Y_{\etap}$ from SU(3)
calculations and experimental branching fractions. 
The horizontal band is the KLOE result assuming $Z_{\etap}=0.$}
\label{bretap}
\end{figure}

This value for $R$ can be related directly to the mixing angle in
the flavor basis. The Bramon {\it et al.}
\cite{BraEsSca99} and Feldmann \cite{Feld00} parametrizations are used
to extract the mixing angle; essentially the same result is obtained 
using both approaches, i.e.,
\fip =$(42.2 \pm 1.7 )^{\circ}$, which gives in a mixing angle in the 
octet-singlet basis $\tp =(-12.9 \pm 1.7 )^{\circ}$.
Moreover, using the value in \cite{PDG} for BR(\fietag), 
 the most precise determination of BR(\fietapg) to date has been extracted 
(see Fig. \ref{bretap}, left):$BR(\fietapg) =\left( 6.1 \pm 0.6 \;({\rm stat.}) \pm 0.4 \;({\rm
syst.})\right) \cdot 10^{-5}$ 

This value for the mixing angle has been obtained neglecting OZI-rule violation and
possible contributions from gluonium to the $\eta$ and $\etap$ mesons.
Allowing for a gluonium component, \cite{Ros83} we write:
\begin{eqnarray}
\ket{\eta}&=&X_\eta\ket{u\bar u+d\bar d}/\sqrt{2}+Y_\eta\ket{s\bar s}+Z_\eta\ket{glue},\nonumber \\
\ket{\eta'}&=&X_{\eta'}\ket{u\bar u+d\bar d}/\sqrt{2}+Y_{\eta'}\ket{s\bar s}+Z_{\eta'}\ket{glue}.
\end{eqnarray}
A non-zero gluonium component of the $\etap$ would correspond to 
$Z_{\etap}^2>0$, or equivalently $X^2_{\etap}+Y^2_{\etap}<$1.
The following constraints on $X_{\etap}$ and  $Y_{\etap}$ can be obtained in a
nearly model-independent way \cite{Ros83,BraEsSca01,Kou00}:
\begin{equation}
\label{X_etap}
\frac{\Gamma(\etap\rightarrow\rho\gamma)}{\Gamma(\omega\rightarrow\pio\gamma)}
\simeq
3\left(\frac{m^2_{\etap}-m^2_{\rho}}{m^2_{\omega}-m^2_{\pi}}\frac{m_{\omega}}{m_{\etap}}\right)^3
X_{\etap}^2
\end{equation}
and
\begin{equation}
\label{X_Y_etap}
\frac{\Gamma(\etap\rightarrow\gamma\gamma)}{\Gamma(\pio\rightarrow\gamma\gamma)}
=
\frac{1}{9}\left(\frac{m_{\etap}}{m_{\pio}}\right)^3(5X_{\etap}+\sqrt{2}Y_{\etap}\frac{f_{\pi}}{f_s})^2.
\end{equation}
The consistency of the assumption of $\eta -\eta'$ mixing without gluonium can
be checked as follows: if $Z_{\etap}=0$, one has $|Y_{\etap}|=\cos{\fip}$.  
This remains a reasonable approximation if the gluonium component is small.
In Fig.~\ref{bretap} (right), the allowed bands corresponding to equations 
(\ref{X_etap}) and (\ref{X_Y_etap}) and the KLOE measurement of $\cos{\fip}$ 
are plotted in the $X_{\etap}$, $Y_{\etap}$ plane, as well as the circumference
$X_{\etap}^2+Y_{\etap}^2=1$, corresponding to zero gluonium in the $\etap$. 
$Z_{\etap}^2$ is seen to be $0.06^{+0.09}_{-0.06}$, which is compatible with 
zero within 1$\sigma$ and consistent with a gluonium fraction of less than 
15\%.
\section{Conclusions}
The data coming from radiative $\phi$ decays are fundamental in
clarifying the nature of the scalar mesons. The branching ratios for $\phi$\to\popog\ and
\phiepg\ and the \fo\ and \ao\ coupling constants have been measured with
the best accuracy by the KLOE collaboration and are in agreement with CMD-2 and
SND results. There is still much to do in this field and more data are
expected from KLOE and from other experiments ($D$ decays etc).
In the pseudoscalar sector there is a new, precise measurement of
BR(\fietapg) and  the \Eta-\etap mixing angle. 
%==============================================================================
% Bibliography
%==============================================================================
%
%\footnotesize

%
%%%%%%%%%%%%%%%%%%%%%%%%%%%%%%%%%%%%%%%%%%%%%%%%%%%%%%%%%%%%%%%%%%%%%%%%%%%%%%%

\begin{thebibliography}{99}
\bibitem{E791_fo} E.M.~Aitala {\it et al}, Phys.\ Rev.\ Lett.\ {\bf86}, 765 (2001).
\bibitem{BNL852} S.~Teige {\it et al}, Phys. Rev. D {\bf 59}, 12001 (2001).
\bibitem{crystal} C.~Amsler {\it et al}, Phys. Lett. B {\bf 333}, 277 (1994). D.V~.Bugg {\it et al},Phys. Rev. D {\bf 50}, 4412 (1994).
\bibitem{4quarks} R.L.~Jaffe, Phys. Rev. D {\bf 15}, 267 (1977).
\bibitem{kk} J.~Weinstein, N.~Isgur, Phys. Rev. Lett. {\bf 48}, 659 (1982).
 N.N.~Achasov, V.V.~Gubin, V.I.~Shevchenko, Phys. Rev. D {\bf 56} 203 (1997).
\bibitem{closetorquist}
  N.A.\,T\"{o}rnqvist,  Hep-ph/0204215 (2002). F.E.~ Close, 
N.A.\,T\"{o}rnqvist,  Hep-ph/0204205 (2002).
\bibitem{kloeao} M.\,Adinolfi {\it et al}, Phys. Lett. B{\bf 538}, 21-26
  (2002).
\bibitem{kloefo} M.\,Adinolfi {\it et al}, Phys. Lett. B{\bf 537}, 21
  (2002).
\bibitem{dafne}
  S.~Guiducci {\it et al},in: Proc.~of the 2001 Particle Accelerator Conference 
  ,(ed. P.\,Lucas S.\,Webber), 353 (2001), (Chicago, Illinois, USA)
\bibitem{sndao} M.N.~Achasov {\it et al}, Phys. Lett. B {\bf 479} 53 (2000).
\bibitem{cmd2ao} R.R.~Akhmetshin {\it et al}, Phys. Lett. B {\bf 462} 380
  (1999).
\bibitem{sndfo} M.N.~Achasov {\it et al}, Phys.\ Lett.\ B {\bf 485} 349 (2000).
\bibitem{cmd2fo} R.R.~Akhmetshin {\it et al}, Phys.\ Lett.\ B {\bf 462} 380
  (1999).
\bibitem{AchIvanch}
  N.N.~Achasov and V.N.~Ivanchenko, Nucl.\ Phys.\ B {\bf 315} 465 (1989).
\bibitem{AchInt} N.N.~Achasov and V.V.~Gubin, Phys.\ Rev.\ D {\bf63} 094007
  (2001).
\bibitem{E791_sigma}
  E.M.~Aitala {\it et al}, Phys.\ Rev.\ Lett.\ {\bf 86} 770 (2001).
\bibitem{E791_sigma1}
  I.~Bediaga,{\it et al} Hep-ex/020839 (2002).
\bibitem{Ankara} A.~Gokalp and O.~Yilmaz, Phys.\ Rev.\ D {\bf 64} 053017 (2001).
\bibitem{WA102}  D.~Barberis {\it et al}, Phys.\ Lett.\ B {\bf 462} 462 (1999).
  F.E.~Close, A.~Kirk, Phys.\ Lett.\ B {\bf 515} 13 (2001).
\bibitem{Escri} A.~Bramon {\it et al}, Phys.Lett. B{\bf 494}, 221 (2000)
\bibitem{Bramon:2002iw}
A.~Bramon, R.~Escribano, J.~L.~Lucio M, M.~Napsuciale and
G.~Pancheri,
%``Scalar f0(980) and sigma(500) meson exchange in Phi
decays into  pi0 pi0 gamma,''
arXiv:hep-ph/0204339.
\bibitem{Marco} E.Marco et al., Phys.Lett. B470 (1999) 20.
\bibitem{BeMor65} C.~Becchi and G.~Morpurgo  Phys. Rev {\bf 140},
B687 (1965).
\bibitem{Close92} F.~E.~Close ``Pseudoscalar mesons at DA$\Phi$NE'' in ``The
DA$\Phi$NE phyisics handbook'' vol.~II (ed.~L.~Maiani, G.~Pancheri and
N.~Paver, Frascati 1992)
\bibitem{Ros83} J.~L.~Rosner  Phys. Rev. D {\bf 27}, 1101 (1983)
\bibitem{BraEsSca97} A.~Bramon, R.~Escribano and M.~D.~Scadron
  Phys. Lett. B {\bf 403}, 339 (1997)
\bibitem{BraEsSca99}  A.~Bramon, R.~Escribano and M.~D.~Scadron  Eur. Phys
  J. C {\bf 7}, 271 (1999)
\bibitem{BraEsSca01}  A.~Bramon, R.~Escribano and M.~D.~Scadron 
Phys.~Lett. B {\bf 503}, 271 (2001)
\bibitem{Feld00} T.~Feldmann,  Int. Jou. Mod. Phys. A {\bf 15}, 159 (2000)
\bibitem{KaisLeut98} R.~Kaiser and H.~Leutwyler, hep-ph/9806336
\bibitem{EsFre99}
R.~Escribano and J.M.~Frere,
Phys.\ Lett.\  B {\bf 459}, 288 (1999)
\bibitem{FeldKroll98} Th.~Feldmann, P.~Kroll, and B.~Stech, 
Phys. Rev. D{\bf 58}, (1998)
\bibitem{Defazio00} F.~De Fazio and M.~R.~Pennington,  JHEP 0007:051
  (2000) 
\bibitem{kloeetap} M.\,Adinolfi {\it et al}, Phys. Lett. B{\bf 541}, 45-51
  (2002).
\bibitem{SND99}
M.N.~Achasov  {\it et al},
 JETP Lett. {\bf 69}, 97 (1999)
\bibitem{CMD297}
R.R.~Akhmetshin {\it et al},
%``First observation of the decay phi $\to$ eta-prime(958) gamma,''
 Phys.~Lett. B {\bf 415}, 445 (1997).

\bibitem{CMD200}
R.R.~Akhmetshin {\it et al.} ,
%``New measurement of the rare decay Phi $\to$ eta' gamma with CMD-2,''
 Phys.~Lett. B {\bf 473}, 337 (2000)

\bibitem{CMD200b}
R.R.~Akhmetshin {\it et al.},
%``Observation of the Phi $\to$ eta' gamma decay with four charged particles  and photons in the final state,''
 Phys.~Lett. B {\bf 494}, 26 (2000)
\bibitem{Kou00} E.~Kou, Phys.\ Rev.\ D {\bf 63}, 054027 (2001)
\bibitem{PDG} The Particle Data Group (D.~Groom {\it et al.})
  Eur. Phys. Jou. {\bf C15} (2000)
%
\end{thebibliography}
\end{document}
%%%%%%%%%%%%%%%%%%%%%%%%%%%%%%%%%%%%%%%%%%%%%%%%%%%%%%%%%%%%%%%%%%%%%%%%%%%%%%%